# MAIN STEPS FOR FABRICATION OF THE IPHI RFQ

M Painchault, J Gaiffier, C Chauvin, J Martin.
SEA / DAPNIA / CEA, France

*Abstract*

The RFQ of the project IPHI [1] is a 8 meter long, high power, very precise tolerances (0.01 mm on 1 meter long for example) device to accelerate protons. This RFQ is similar to the RFQ of the LEDA project. So, we realize a thermal and mechanical studies followed by different tests for machining and brazing copper. We describe in this paper those different steps and the way we proceed to supply the fabrication itself by an independent company.

## 1 SUBMISSION OF THE PROBLEM

The main difficulties to fabricate this RFQ and consequences on the design are summarized on table 1:

| Main characteristics | Consequences |
|---|---|
| ▪ 1.2MWHF = 15 W/cm$^2$ average with pick at 150 W/cm$^2$.<br>▪ High Q; low ΔF<br>▪ Waved profile of the vane tip<br><br>▪ Good vacuum | ▪ Structure in copper.<br>▪ Control of thermal expansion<br>▪ Drilling the cooling channels.<br>▪ High tolerances:<br>+ Minimum number of pieces<br>+ brazing in 2 steps<br>+ Precise machining<br>▪ Metallic linkage for helicoflex |

Table1: Characteristics

So, we first do a thermal study to determine dimensions and positions of the channels and optimize the cooling characteristics.

Second, we design the section.

Third, we do a lot of tests to qualify the machining and brazing procedures.

## 2 THERMAL STUDY

### 2.1 Cooling the vane.

#### 2.1.1 – Topics

We search a correct frequency of the cavity (+/- 50 kHz), a constant frequency all along the cavity, vanes at the right position, a maximum stress below the yield point.

On another hand, it would be more convenient during running operations to obtain small displacements on the vane tip in transition cold / hot period. So, we want the lowest running temperature as possible.

#### 2.1.2 Method of calculation.

The geometry is given on figure 1:

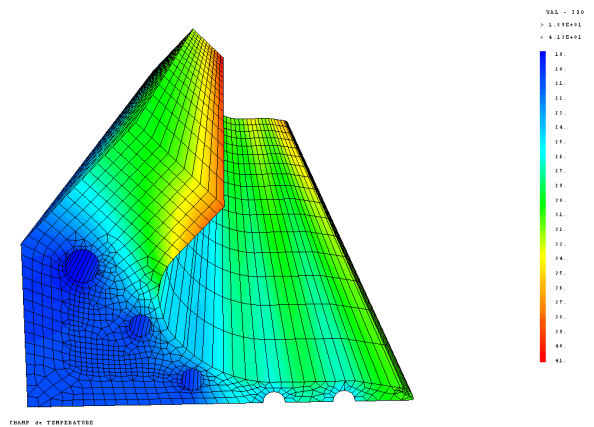

Figure1: temperatures map

First stage is the optimization of the initial 2D 8$^{th}$ section which receives the greatest power density. So we did the following steps:

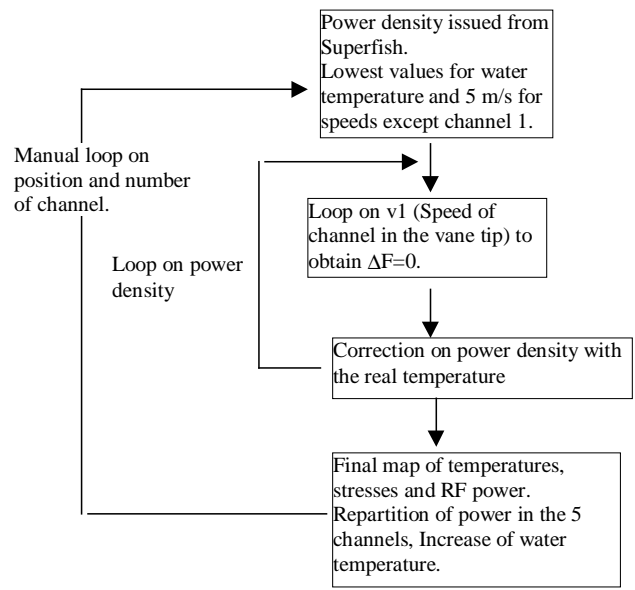

Second stage: 2D calculation of final section: We just verify that there is no important derive.

Third stage: 3D calculation of the 8$^{th}$ sections.

We obtain the temperatures map, stresses map. We have also the evolution of the vane tip position, the

evolution of the bottom of the cavity, the evolution of the frequency all along the section:

Fourth stage: 3D: Inlet temperature of channel 1 and 2 are fixed at 10°C. We search a single inlet temperature of channel 3,4 and 5. In fact the RFQ will be regulate by channels of the bottom of the cavity.

### 2.1.3 Results:

The temperatures map is given on figure 1. Maximum temperature is 41°C on the bottom. The vane tip are at 13°C. Von Mieses stresses were below 26 MPa, frequency varies between 0 and –50kHz without the extremities. Vane tip moves less than 5 μm.

Stresses are greater in 3D than in 2D. So, we re-optimize the geometry.

Inlet temperatures of channel 3,4,5 vary between 10°C and 13°C from section2 (minimum) and section 8 (maximum).

### 2.2 Segment ends

Local RF heat is underestimated by RF simulations. P Balleyguier shows at lease a factor 4 on the power displayed on the edges [2]. But, in fact, global values are reliable. They give a good idea of displacements.

Displacements are given on figure 2 where the maximum is 0,07 mm at the end of the vane type (blue color):

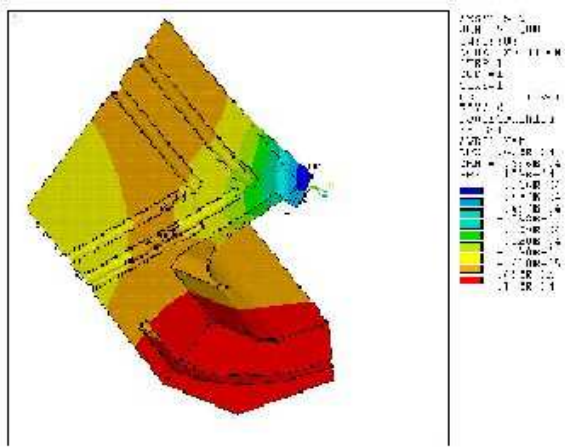

By the way, an inclined end instead of a rectangular cut like LEDA RFQ decreases displacements of the vane tip.

The maximum temperature is evaluated to 80°C. The maximum von Mieses stress is about 80 MPa.

So, the conclusions for the design is to do smooth edges to spread the RF power and to reduce distance between the channel and the end face. This work is still in progress to be more confident on local values and reduce temperature and stresses.

## 3 COPPER

We are chiefly interested by grains size and thermal expansion. For the thermal expansion coefficient and the brazing operation, we need chiefly an homogenous copper in the three directions. Grains size is important to avoid leaks.

Properties of copper are strongly dependent on its history. So, we specify hardness, easier to measure and equally dependent on the way to elaborate it.

The copper specification precises this two parameters.

Of course, we fabricate all copper in a single time. We control the forging at the supplier.

## 4 DESIGN OF THE RFQ

The design lays upon the principles detailed in the first paragraph and upon results of the thermal studies.

Further, to determine the way to cut the RFQ in segment, in section of 1 meter long, each with four vanes: 2 majors and 2 minors, we completely reproduce the LEDA schedule.[3]

We add stainless steel flange for metallic linkage. Flanges for the tuner are also in stainless steel. The output for the cooling is reported on the copper structure.

The two stages of brazing are one for stainless steel components, one for copper pieces.

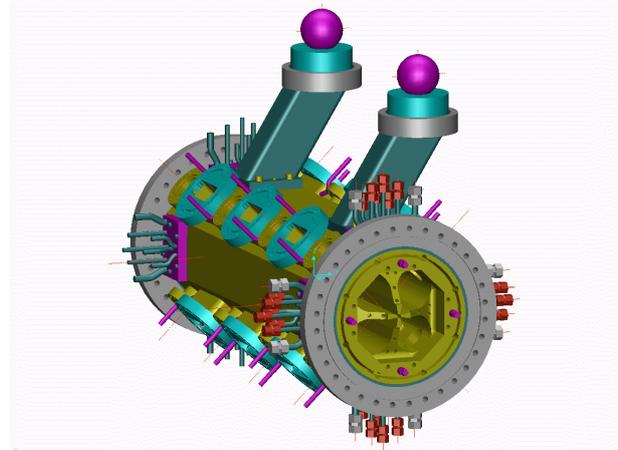

Figure 3: section 1

To align, we will use taylor balls.

The general operations of the fabrication procedure is chronologically:

+ Drilling the channels,
+ Rough machining of the vanes,
+ Final machining of the major vanes and control,
+ Final machining of the minor vanes and control,
+ Surface treatment,
+ Assembling minor and major vanes,
+ RF tests,
+ Last machining issued from the assemble,
+ Brazing all components together and control (dimensions and RF).

# 5 THE DIFFERENT STEPS OF DEVELOPMENT.

We have two topics: first determine the best way for machining and brazing (it is the technical topic), second to find a supplier for these operations (the commercial purpose).So, we plan the following schedule:

| Technical steps | Topics | Commercial steps | Calendar |
|---|---|---|---|
| 1 – Preliminary tests | = on small pieces. | Invitation to tender | 3 months |
| 2 – representative rough model. (500 mm long) | + analyse the manufacturing constraints + First procedure | Invitation to tender | 3 months |
| 3 – Tests on 1 meter long | To validate all parameters | First step of the market: firm | 10 months |
| 4 – pre – prototype | First repetition | First step of the market: firm | 3 months |
| 5 – Prototype | | Second step of the market: firm | 3 months |
| 6 – All sections. | | Optional: depends on results on the prototype | 20 months |

Table2: schedule of development.

To design the rough model, we first design the prototype and we keep the most sensible aspects.

# 6 ROUGH MODEL.

It is 500 mm long, has a single vane, one assembled face, and reference faces to see deformations.
Results are: Form tolerance of the tip vane: 0,02 mm, flatness of assembled faces: 0,01 mm, parallelism between the assembled face and the vane tip: 0,01 mm. The brazing operation add a 0,01 mm deformation. The drilling test gives an displacement below 0,5 mm but the operation stays risky.
It was the hoped values which were held by SICN. So, it was the chosen supplier.

# 7 TEST ON 1 METER LONG.

In fact, we identify the technical difficulties and imagine a test for each one. Those tests and their results are given on table 3.
One of them is the assemblage of stainless steel and copper. We chose a brazing solution because electron beam welding has been tested to assemble copper with too greatest deformations. The brazed method needs precise machining tolerances but experience of precise values were obtained for vanes.
To determine the desired tolerance, we use a thermal mechanical program called Castem, developed by CEA. This program can simulate the elastic and plastic behavior of materials during thermal cycles.

| Tests | Topics | Results |
|---|---|---|
| Machining faces on 1 m long block + thermal treatment. | To verify possibilities of the machine and the effect of high temperatures. | Flatness 0,01 mm. Variation 0,02 mm. |
| Drilling 1 m long holes. | Feasibility | Precision 0,4 mm |
| Material control after thermal treatment | Characterize grain size. | Done. |
| Brazing stainless steel flange on copper cavity | To minimize deformation on the vane tip. | Variation of the position below 0,01 mm. |
| Brazing on 500 mm long | To verify the brazing conditions | Deformation: 0,01 mm ; Compactness with ultrasounds: 90% |
| Brazing on 1m long | To qualify the whole procedure before the pre prototype. | In progress |
| Obtrusion of holes | To qualify the procedure | No leak |
| Machining and brazing the pumping ports | To qualify the procedure before the pre prototype | Tolerance within 0,01 mm |
| Machining the definitive vane tip on 300 mm long and control it. | To validate the machining program and determine precision of tools. To validate the control procedure. | Form factor < 0,02 mm. |
| Machining the final vane tip on 1 m long. | To validate on the true dimensions | In progress |

# 8 PRE - PROTOTYPE.

The pre – prototype is mechanically the same than the prototype but we simplify it to reduce costs. For example, there is only one vane tip machined with the correct profile, holes for pumping ports are done but just one will be brazed, few cooling holes will be drilled.
At the opposite, we could make cuts to visualize deformations and see the quality of the brazed junction.
We yet have roughly machined the pieces. We still have to machine the waved profile on one major piece and to braze the pre - prototype.

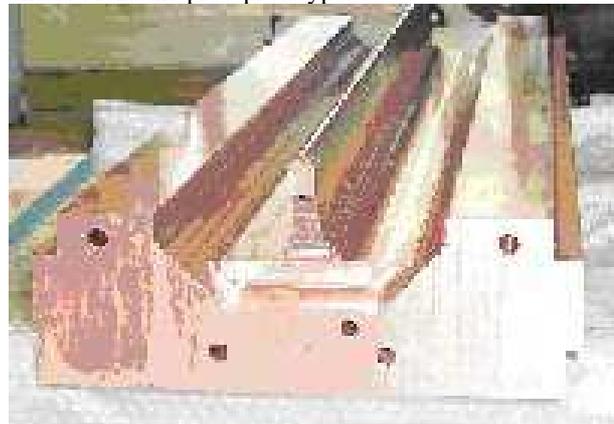

Figure 4: major vane roughly machined

# 9 CONCLUSION.

We plan to achieve the pre prototype for middle of October 2000, fabrication of the first section would be ended beginning 2001 and the RFQ in middle of 2002.

## REFERENCES

[1] R Ferdinand, Status report on 5 MeV IPHI RFQ, linac 2000.
[2] P Balleyguier, 3D design of the IPHI RFQ cavity, this conference (linac 2000, Monterey).
[3] D L Schrage, CW RFQ Fabrication and Engeenering, Linac 1998